\documentclass[reprint,superscriptaddress, amsmath,amssymb, aps, prl, ]{revtex4-2}

\usepackage[usenames, dvipsnames]{color}
\usepackage{comment}
\usepackage{hyperref}
\usepackage{subcaption}
\usepackage{float}
\usepackage{ragged2e}
\hypersetup{
    colorlinks=true,
    linkcolor=Blue,
    citecolor=Blue,
    filecolor=Blue,      
    urlcolor=Blue,
    pdftitle={Universal Attractive and Repulsive Branches of 1D Bose Polarons}}
\usepackage{graphicx}
\usepackage{physics}
\usepackage{xcolor}

\begin{document}
\title{Unified theory of attractive and repulsive polarons in one-dimensional Bose gas}
\author{Nikolay Yegovtsev}
 \email{nikolay.yegovtsev@colorado.edu}
 \affiliation{%
Department of Physics and Astronomy and IQ Initiative,
University of Pittsburgh, Pittsburgh, Pennsylvania 15260, USA\\
}%
\author{T. Alper Yo\u{g}urt}
 \email{ayogurt@pks.mpg.de}
 \affiliation{%
Max Planck Institute for the Physics of Complex Systems, N\"othnitzer Str. 38, 01187 Dresden, Germany\\
}%
\author{Matthew T. Eiles}
 \affiliation{%
Max Planck Institute for the Physics of Complex Systems, N\"othnitzer Str. 38, 01187 Dresden, Germany\\
}%
\author{Victor Gurarie}
 \affiliation{%
Department of Physics and Center for Theory of Quantum Matter,
University of Colorado, Boulder Colorado 80309, USA\\
}%
\date{\today}

\begin{abstract}
    We present a unified description of attractive and repulsive polarons, formed in a one-dimensional Bose gas hosting an impurity particle, by obtaining all ground and excited state solutions to the Gross-Pitaevskii equation. 
    Modeling the impurity with an attractive square-well potential, we characterize the excited-state energy branches as a function of interaction strength.
    As the impurity-bath coupling increases, the excited states change from distinct soliton configurations to hybridized soliton-polaron states, eventually crossing over from repulsive to attractive polarons at unitarity. 
    We identify a universal regime near this crossover
    where the polaron properties are accurately characterized by the zero-energy scattering length.     
\end{abstract}

\maketitle

\textit{Introduction--} The study of quantum impurities immersed in a complex environment has attracted significant attention in the atomic physics community in recent years \cite{massignan2025polarons, grusdt2025impurities,scazza2022repulsive} due to 
the tunability of impurity–bath interactions and the high level of control provided by ultracold atoms, which allow one to study new physical phenomena that are beyond the reach of standard solid-state materials. One particular direction of current research is to understand the effects of the environment on the impurity within the Landau quasiparticle picture -- the so-called polaron problem \cite{Landau:1933iwn} --across both weak and strong bath-impurity coupling regimes. This is motivated by the attempt to develop new tools for understanding strongly correlated phenomena in real materials. Moreover, recent studies showed that impurities immersed in a complex environment can inherit its topological properties \cite{grusdt2016interferometric, PhysRevB.103.245106, PhysRevB.104.035133,grusdt2019topological, PhysRevB.99.081105,2020_PRL_Lewenstein_Anyon_Impurity, vashisht2025chiral} or exert significant influence on it \cite{jager2020strong, Massignan2021, PhysRevResearch.6.L022014}, suggesting that they could, in principle, serve as novel tools for measurement and control. 

An impurity in a one-dimensional (1D) Bose gas presents a good example of both the polaronic and topological aspects of this problem, as its description requires theoretical approaches beyond the conventional polaron theories in the strong coupling regime \cite{1D_Polaron_Experiment,grusdt2017bose,2017_PRA_Volosniev_1D_Polaron,2020_PRR_Peter_Induced,will2021polaron,petkovic2103mediated,2025_PRA_Micheal_Impurity_Localization,yougurt2025analytical} while its coupling to the environment provides a means to manipulate solitons via external control over the impurity \cite{2002_PRA_Peter_Dark_Soliton_Impurity_Interactions,2016_PRA_Axel_Impurity_1D_BEC,2019_PRA_Peter_Impurity_Soliton,mostaan2022quantized, meng2024controlling, PhysRevResearch.6.L032040,2024_PRA_Cao_Soliton_Impurity_Pumping,majumdar2025relaxation}. Although much effort has been put into understanding polarons in the dilute bath limit, where the bath-impurity interaction can be faithfully modeled by a contact potential, with the recent development in the trapping of ions and Rydberg atoms inside the Bose condensate or degenerate Fermi gases \cite{PhysRevX.6.031020, zipkes2010trapped, RevModPhys.91.035001, PhysRevLett.129.153401}, the effects of a finite bath-impurity potential range can come into play and lead to novel physical effects \cite{christensen2021charged,sous2020rydberg,camargo2018creation,yougurt2025quasiparticle,durst2024phenomenology, yegovtsev2025fermipolaronsfiniterangefermionimpurity}. 

In this Letter, we consider an infinitely heavy impurity immersed in a one-dimensional Bose gas and interacting with it via a short-ranged potential of finite extent $r_c$. Our theoretical formalism is based on the Gross–Pitaevskii equation (GPe), well known as the appropriate tool to study 1D Bose gas at sufficiently weak interactions among bosons \cite{1963_Lieb_Exact,1963_Lieb_Exact_2,jager2020strong,will2021polaron}. The impurity is treated as an external potential acting on the weakly interacting Bose particles. 
The GPe approach accounts for the substantial condensate distortions induced by the impurity and is necessary for an accurate description of polarons in the strong-coupling regime~\cite{jager2020strong, Massignan2005, PhysRevA.110.023310}. We show that finite range effects are essential for understanding the spectrum of excited states. 
Because of the high compressibility of the Bose bath, the effect of a single impurity is significant: the impurity becomes strongly coupled to single- or multi-soliton configurations, supported by the gas in the absence of the impurity-- 
and near unitary points of the boson-impurity potential, the energy of these configurations has a universal polaron-like description. Although we present our findings using an attractive square-well potential to make the problem analytically tractable and simplify the numerical analysis, our results apply to arbitrary short-ranged potentials. 
The interaction of an impurity with solitons has been partially addressed in \cite{2002_PRA_Peter_Dark_Soliton_Impurity_Interactions, 2016_PRA_Axel_Impurity_1D_BEC, 2015_PRL_Polaron_to_Soliton, 2025_Arxiv_Impurity_Relaxation_Dynamics_Solitons, petkovic2016dynamics, will2023dynamics, PhysRevA.85.023623, grusdt2017bose, hakim1997nonlinear, koutentakis2021pattern}; however, a comprehensive investigation of all possible excited state solutions, their relation to the polaronic states, and their universality has been missing until now.

\begin{figure*}[t]
        \includegraphics[width =1\textwidth,keepaspectratio]{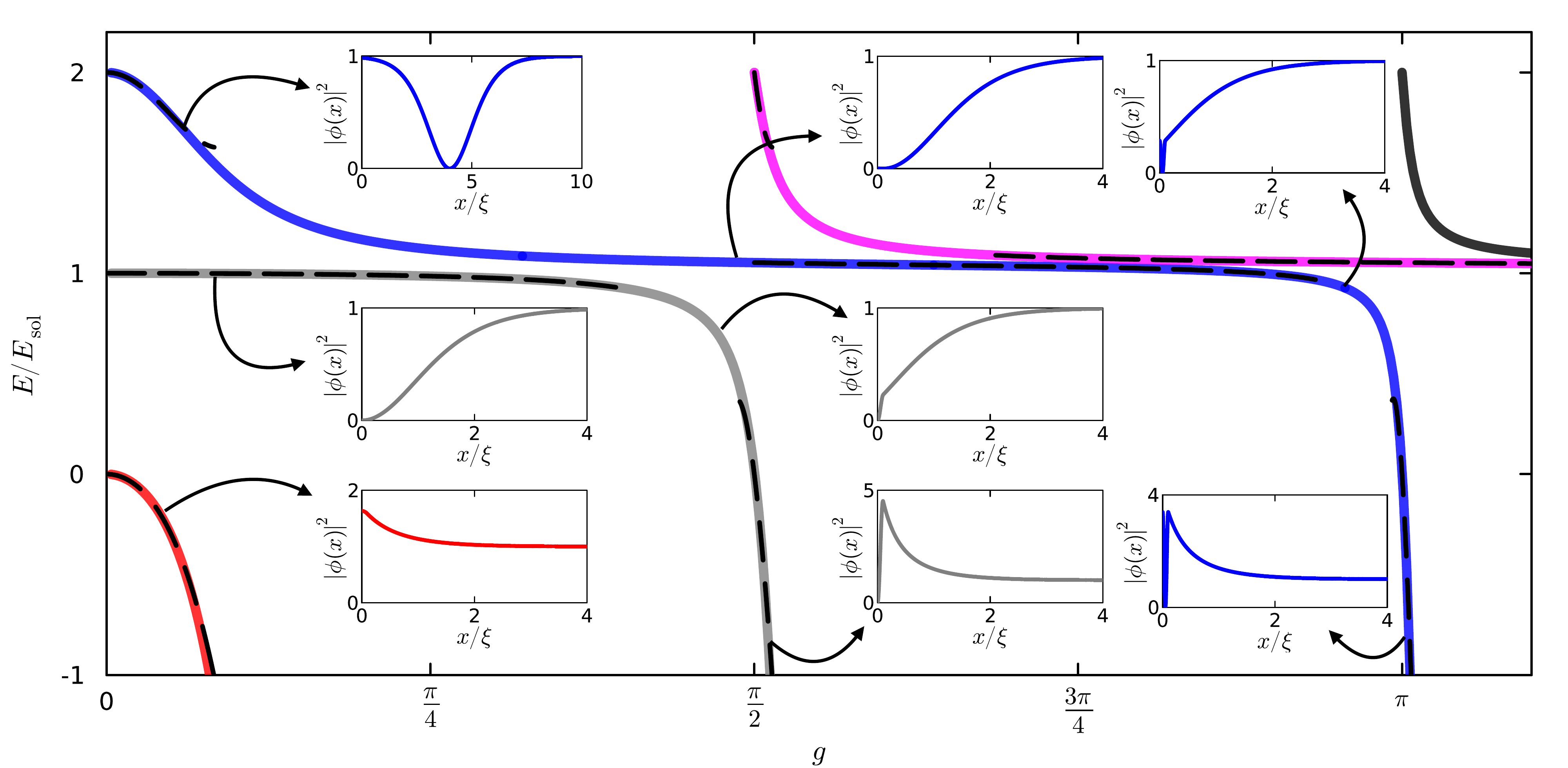}
    \caption{\justifying Energy branches for an attractive square well potential as a function of coupling constant $g$. The insets display the density profiles $|\phi(x>0)|^2$ for the evolution of the ground-state (red solid), first odd-parity excited state (gray solid), and first even-parity excited state (blue solid) solutions to the GPe~\eqref{eq:LGPe0inText}. The dashed lines indicate the regimes where the energy of the solution can be described by the universal expansions for polaronic Eq.~\eqref{eq:dpwc} or solitonic Eq.~\eqref{eq:dpfc} configurations with the correct choice of $a_s$ and $a_p$ in Eq.~\eqref{eq:asap} for the even or odd states.
    }
    \label{fig:Branch_wavefunctions}
\end{figure*}

In this work, we focus only on real solutions to the GPe that are regular. Our main result is presented in Fig.~\ref{fig:Branch_wavefunctions}, where we show all physically admissible solutions to the GPe as a function of the strength of the impurity potential and show that the energy of such solutions in polaronic and solitonic regimes can be described by universal expansions given in Eq.~\eqref{eq:dpwc} and Eq.~\eqref{eq:dpfc} involving zero-energy scattering length corresponding to the parity of a given state.

\textit{Theory--} The 1D GPe for an infinitely heavy impurity reads ($\hbar=1$)
\begin{equation}
\label{eq:GPl} \left(- \frac{\partial^2_x }{2m} + U(x)  -  \mu  + \lambda | \psi(x) |^2 \right)  \psi(x) = 0,    
\end{equation}
where $\psi(x)$ is the boson field, $m$ is the mass of the bosons, $\lambda$ is the strength of the boson-boson repulsion, and the chemical potential $\mu = \lambda n_0$ is set to ensure a uniform density $n_0$ far away from the impurity. 
We consider a symmetric short-ranged impurity-bath interaction $U(x)$ that vanishes beyond some characteristic distance $r_c$, which we set as the unit of distance. 
Introducing the dimensionless parameter $\epsilon \equiv r_c/\xi$, where $\xi = 1/ \sqrt{2m\lambda n_0}$ is the coherence length, leads to the GPe
\begin{equation}
-\partial^2_y\phi + \mathcal{U}(y)\phi  =\epsilon^2\left(\phi-\phi^3 \right),
\label{eq:GPe0Text}
\end{equation}
where $y \equiv x/r_c$, $\phi \equiv \psi/\sqrt{n_0}$ and $\mathcal{U}(y)\equiv 2mr_c^2U(y)$.
The 
energy of the state $\phi$, relative to the uniform gas, is \cite{Massignan2021,Yegovtsev2022}
\begin{equation}
\label{eq:penegytext}
E = -\frac{\lambda n_0^2}{2}\int_{-\infty}^{+\infty}dx \left(\phi^4-1 \right). 
\end{equation}
 We set $\lim_{y\to+\infty}\phi(y)=1$ without loss of generality. 
The solution of Eq.~\eqref{eq:GPe0Text} for $U(y)=0$ is given by
\begin{equation}
\label{eq:dpphiText}
\phi(x) = 1 - \frac{2c}{c-2e^{\sqrt{2}x/\xi}}.   \end{equation}
Here, the parameter $c$ specifies the density profile outside the impurity potential range. Specifically, Eq.~\ref{eq:dpphiText} yields the dark soliton profile $\tanh(x/(\sqrt{2}\xi))$ when $c=-2$, and corresponds to a flat density at $c=0$.

\begin{figure}[t]
    \begin{subfigure}{0.47\textwidth}
        \includegraphics[width = \textwidth,keepaspectratio]{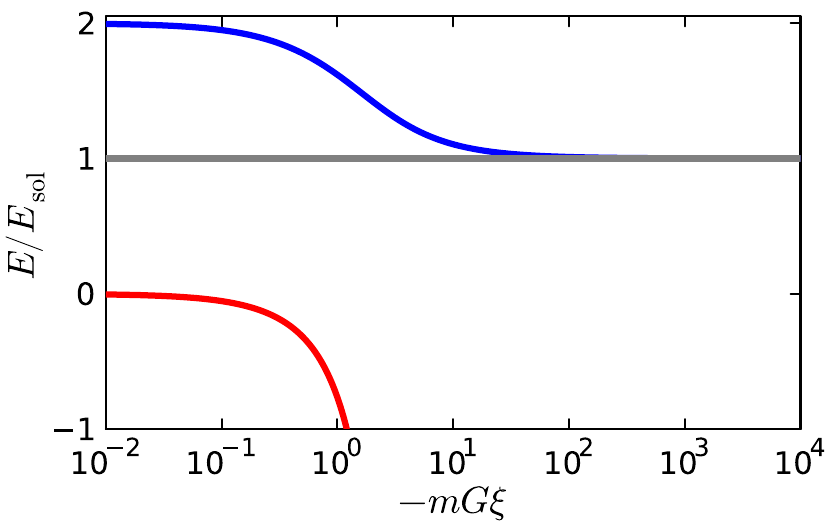}
    \end{subfigure}
    \caption{\justifying Energy branches for an attractive contact impurity, given as a function of the coupling strength $mG\xi$ and obtained for the ground-state (red) and excited-state (gray and blue) solutions of the Gross–Pitaevskii equation Eq.~\eqref{eq:GPl}. 
    }
    \label{fig:Branch_Figure}
\end{figure}

\textit{Contact impurity potential--} To motivate our study of finite-range boson-impurity potentials, let us briefly consider a contact impurity potential
\begin{equation}
\label{eq:Ucontact}
U(x) = G\delta(x),    
\end{equation}
to illustrate its limitations. 
Prior studies have focused exclusively on ground-state solutions, addressing both $G<0$ and $G>0$ scenarios \cite{grusdt2017bose, jager2020strong, PhysRevA.105.L021303}; here we focus on $G<0$, which supports only one bound state regardless of the magnitude $|G|$. In this regime, we find three distinct solutions to Eq.~\eqref{eq:GPl}. At $G=0$, the first excited-state solution corresponds to the dark soliton (gray curve in Fig.~\ref{fig:Branch_Figure}). Because of the odd parity, this solution remains unaffected by the presence of the contact impurity as $|G|$ increases, and its energy remains constant
$$E_{\mathrm{sol}} =  \frac{2\sqrt{2}}{3}\frac{n_0}{m\xi}.$$ 

The remaining two solutions exhibit even parity and are obtained for 
\begin{equation}
\label{eq:cdeltatext}
c_{\pm} = \frac{2(\sqrt{2} \pm \sqrt{2+(mG\xi)^2})}{mG\xi}.   
\end{equation}
Here, the $c_-$ branch describes the ground-state solution -- the attractive polaron -- characterized by an increased density of atoms in the vicinity of the impurity. 
The energy of this branch (Red curve in Fig.~\ref{fig:Branch_Figure}) can be computed from Eq.~\eqref{eq:cdeltatext}; by expanding the result around $G=0$ or $G = - \infty$ and expressing the result in terms of the scattering length $a=-1/(mG)$, we obtain 
\begin{equation}
\label{eq:dpwc}  E  = \left\{ \begin{matrix}   -\frac{n_0}{m\xi}\left[\frac{\xi}{a} + \frac{1}{\sqrt{2}}\left(\frac{\xi}{a}\right)^2 + \mathcal{O}[(\xi/a)^3]\right], & \frac{\xi}{a}\ll1 , \cr -\frac{n_0}{m\xi}\left[\frac{2}{3}\left(\frac{\xi}{a}\right)^3+2\frac{\xi}{a}+\mathcal{O}(a/\xi)\right],  &\frac{a}{\xi}\ll1. 
\end{matrix}
\right.    
\end{equation} 
From this, we see that the small parameter of the theory is given by $\xi/a$ in the weak-coupling or $a/\xi$ in the strong-coupling regime; this connection to scattering length allows us to use this expansion for finite-range potentials in the remainder of the paper.
Eq.~\eqref{eq:dpwc} shows that the ground-state energy is adiabatically connected to that of the uniform condensate in the limit $G \to 0$.

The other excited-state solution (with $c_+$) describes a soliton-antisoliton
pair spaced infinitely far apart as $G\to 0$.  
As $|G|$ increases, the centers of both soliton and antisoliton move toward the origin, and in the limiting case $|G|\to \infty$ they merge, resulting in a configuration $\phi = \tanh(|x|/(\sqrt{2}\xi))$ (Blue curve in Fig.~\ref{fig:Branch_Figure}). This has the same density profile as a single dark-soliton. Note that although the impurity can significantly affect the solution of the $c_{+}$ branch, the energy scale always remains in the solitonic regime and does not transition into the polaronic regime described by Eq.~\eqref{eq:dpwc}. 
The energy expansion for this branch is
\begin{equation}
\label{eq:dpfc}  E  = \left\{ \begin{matrix}   2E_\mathrm{sol}-\frac{n_0}{m\xi}\left[  \frac{\xi}{a} - \frac{1}{\sqrt{2}}\left(\frac{\xi}{a}\right)^2 + \mathcal{O}[(\xi/a)^3]\right], & \frac{\xi}{a}\ll1 , \cr E_\mathrm{sol}+\frac{n_0}{2m\xi}\left[\frac{a}{\xi} +  \mathcal{O}[(a/{\xi})^3]\right],  &\frac{a}{\xi}\ll1, 
\end{matrix}
\right.    
\end{equation} from which we see the adiabatic transition from an antisoliton-soliton to a single soliton-like configuration as the interaction strength increases.  

\textit{Square well impurity potential--} The above analysis shows that the contact impurity potential has several key limitations. First, the presence of the contact impurity does not affect the odd-parity states. Second, the contact impurity potential 
can support at most a single bound state, so the number of possible solutions to the GPe satisfying Eq.~\eqref{eq:dpphiText} outside of the impurity potential was limited to three.
A similar problem has been studied in 3D, where it was shown that the GPe can yield up to $2\nu_b+1$ spherically symmetric solutions when the impurity-bath interaction potential supports $\nu_b$ bound states  \cite{Massignan2005}. Although Ref.~\cite{mostaan2023unifiedtheorystrongcoupling} recently studied the existence of other excited states in 3D, a complete characterization of all excited state branches and the possibility of a universal description has not yet been carried out. While this remains an outstanding problem in 3D, here we show that it can be completely resolved in 1D. We focus our analysis on the case of an attractive square well $\mathcal{U}(y)=-g^2\Theta(1-y)$, which yields analytical solutions to Eq.~\eqref{eq:GPe0Text}. The GPe inside the impurity range becomes
\begin{equation}
\label{eq:LGPe0inText}
-\partial^2_y\phi -g^2\phi  =\epsilon^2\left(\phi-\phi^3 \right).    
\end{equation}
The even and odd-parity solutions are given, respectively, by 
\begin{equation}
\label{eq:cdText}
\phi_s = \frac{1}{\epsilon}\sqrt{\frac{2c_s(g^2+\epsilon^2)}{1+c_s}}\text{cd}\left(\sqrt{\frac{g^2+\epsilon^2}{1+c_s}}y,c_s\right),  
\end{equation}
\begin{equation}
\label{eq:snText}
\phi_p = \frac{1}{\epsilon}\sqrt{\frac{2c_p(g^2+\epsilon^2)}{1+c_p}}\text{sn}\left(\sqrt{\frac{g^2+\epsilon^2}{1+c_p}}y,c_p\right),  
\end{equation}
where $\text{cd}(y,c_s)$  and $\text{sn}(y,c_p)$ are Jacobi elliptic functions \cite{1948_Abramowitz_Mathematics,2013_Lawden_Elliptic_Functions}, with elliptic parameters $c_s,c_p\in [0,1]$. Outside the impurity region, the solution is still given by Eq.~\eqref{eq:dpphiText}. The unknowns $c$, $c_s$, and $c_p$ are determined by the boundary conditions at $y=1$ \footnote{Note that while we fixed the behavior of the solution at $y\to \infty$ of the tail in \eqref{eq:dpphiText}, the solutions inside the impurity potential still possess a $Z_2$ symmetry, so when we numerically match \eqref{eq:cdText} with \eqref{eq:dpphiText} or \eqref{eq:snText} with \eqref{eq:dpphiText} and their derivatives, we also need to consider the case when $\phi_s\to-\phi_s$ and $\phi_p\to -\phi_p$ to obtain all possible solutions.}. 

We are interested in the regime $r_c\ll\xi$, where the polaron's properties depend primarily on the scattering length. For the attractive square well potential in 1D we have two scattering lengths $a_s$ and $a_p$, corresponding to even and odd parity states
\begin{equation}
\label{eq:asap}
\begin{split}
&a_s = r_c\left(1 + \frac{\cot g}{g}\right),\\
&a_p = r_c\left(1 - \frac{\tan g}{g}\right).
\end{split}
\end{equation}
The scattering length diverges whenever a new bound state appears in the spectrum. This occurs at $g = n\pi$ for even states and $g = (n+1/2)\pi$ for the odd states, with integer $n \geq 0$. 

\textit{Results-} 
Having established the necessary theoretical framework, we now present the numerical analysis of the admissible solutions (energy branches) to the GPe. We fix $\epsilon = 0.1$ to model a short-range impurity potential and vary $g$ from 0 to $2\pi$. 
In general, we find $\nu_s+\nu_p+2$ number of distinct solutions, where $\nu_s$ and $\nu_p$ are the number of even and odd bound states supported by $U(x)$, respectively. 
Each of these energy branches is depicted using a distinct color in Fig.~\ref{fig:Branch_wavefunctions}. In the vicinity of $g=0$ these branches behave similarly as in the case of the contact potential; however, new effects emerge as we start increasing $g$. Near unitarity in the relevant parity channel, 
the excited state branches curve rapidly away from the contact potential's asymptotic values, and indeed even become negative. New excited-state branches appear near the unitary points, where one of the roots of the matching equation produces $c<2e^{\sqrt{2}\epsilon}$ that results in a regular behavior of the tail in Eq.~\eqref{eq:dpphiText}. Physically, these new branches around $g = n\pi/2$ can be regarded as originating from a configuration of $n+2$ solitons that are infinitely separated at $g=0$. This correspondence is further reflected in the equivalence between the number of nodes in the emerging bound state and the number of solitons in the zero-coupling limit. As the impurity potential deepens, the solitons gradually approach each other and eventually become resolvable within our method, which relies on the form of the tail in Eq.~\eqref{eq:dpphiText}. 


We observe that near the unitary points, the solution profiles of the GPe inside the impurity potential can be well approximated by the solution to the zero-energy Schr\"odinger equation. This suggests that, provided that $\epsilon \ll 1$ when we match the interior and exterior solutions at $y=1$, we can simply impose a Bethe–Peierls boundary condition at the edge of the potential. This calculation is analogous to the one done in \cite{Massignan2021, Yegovtsev2022} in 3D, so we will not reproduce it here. We just note that, unlike in 3D, where the potential's finite range was essential for deriving a new universal expansion around a unitary point, in 1D the resulting calculation reproduces the contact potential result given in Eq.~\eqref{eq:dpwc} and Eq.~\eqref{eq:dpfc}.
Even and odd parity state energies are obtained by substituting $a_s$ or $a_p$, respectively, for the scattering length 
\footnote{The applicability condition of these expansions for an arbitrary finite-ranged potential can be derived by demanding that the contribution to the energy from the interior of the well is much smaller than the contribution from the tail of the polaron, and reads $r_c/a\ll1$. Once $a\sim r_c$, we start to see other details of the potential that are not universal. }.
\begin{figure}[t]
        \includegraphics[width =0.48\textwidth,keepaspectratio]{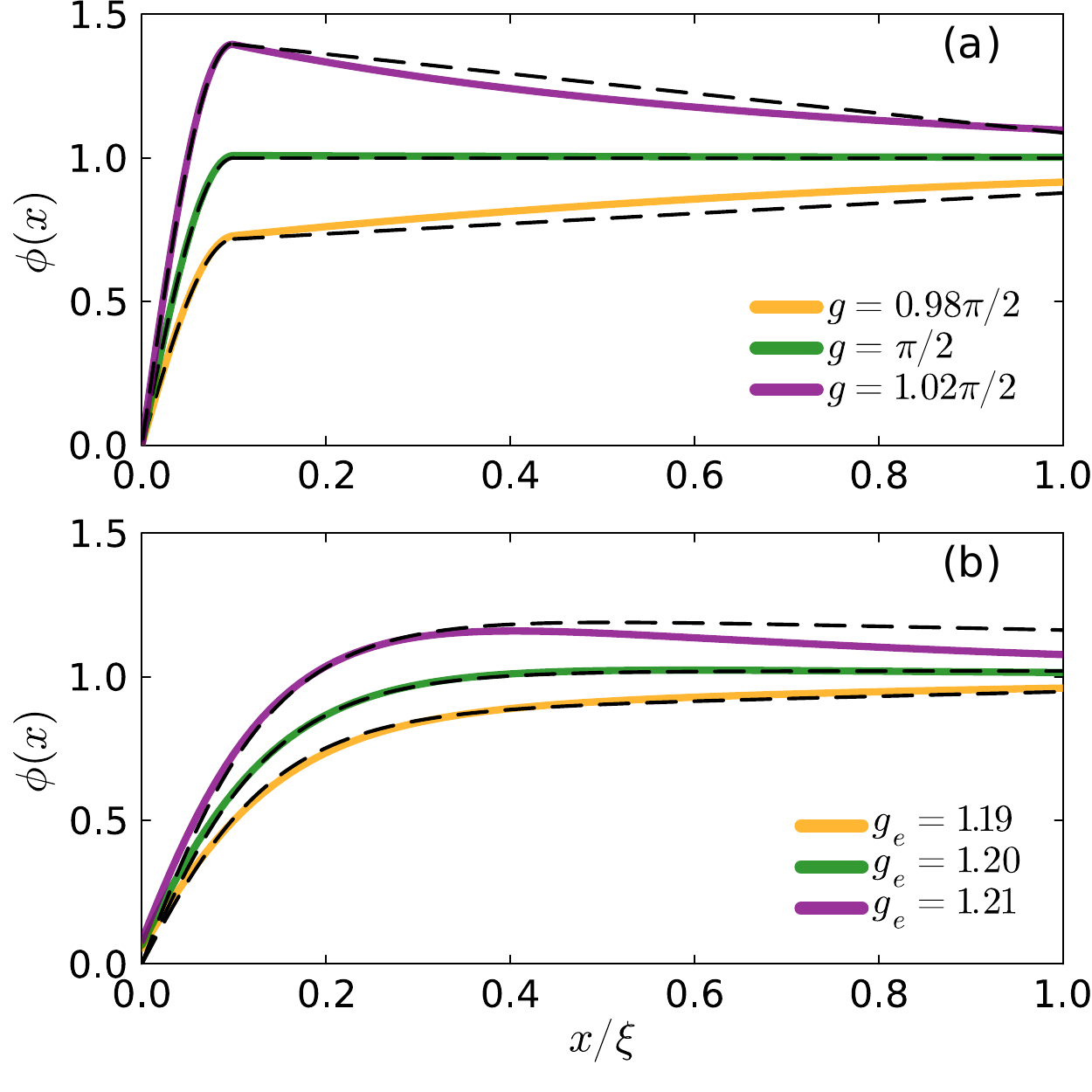}
    \caption{\justifying Repulsive-to-attractive polaron crossover near the first unitary point of the odd-parity excited state for an impurity potential of \textbf{(a)} a square well and of \textbf{(b)} an exponential potential. Solid lines show the exact solutions of the GPe in Eq.~\eqref{eq:LGPe0inText}, while dashed lines correspond to solutions 
    of the zero-energy Schr\"odinger equation.
    }
    \label{fig:transition_near_unitarity}
\end{figure}

To further illustrate the qualitative evolution of the solutions along each branch, representative wavefunctions for the ground state (red curve) and the first two excited states (gray and blue curves) are shown in Fig.~\ref{fig:Branch_wavefunctions} as insets. These insets are selected to highlight the hybridization between the soliton configurations and polaronic states, and the repulsive-to-attractive polaron crossover. Universal expansions in Eq.~\eqref{eq:dpwc} and Eq.~\eqref{eq:dpfc} together with the analytic expressions for $a_s$ and $a_p$ in Eq.~\eqref{eq:asap} allow us to interpret the structure of these energy branches. First, we comment on the evolution of the ground state branch. It is described by Eq.~\eqref{eq:dpwc} in the regime of weak and strong-coupling expansion. When $a_s\sim r_c$, we observe numerically 
the solution starts to scale as $g/\epsilon$ and is no longer universal. When a new excited state energy branch (apart from the first odd branch for which $a_p \leq r_c$ from the beginning) appears near a corresponding unitary point, it is described by Eq.~\eqref{eq:dpfc}. Then, as the scattering length decreases in magnitude and becomes comparable to $r_c$, we expect the universal description to break down. Eventually, as we approach a new unitary point of the same parity sector, this solution that had a soliton-like structure continuously evolves into the polaronic solution described by Eq.~\eqref{eq:dpwc}. Interestingly, since the energy of such a configuration is positive, the resultant polaron configuration is the repulsive polaron until the impurity potential crosses a unitary point. As it crosses the unitarity, the energy of the configuration becomes negative, and we call this transition a repulsive-to-attractive polaron crossover. Once the energy of such a polaron becomes sufficiently negative, the microscopic details of the potential start to come into play, similar to the case of the ground state branch.
The corresponding energy expansions, shown as dashed lines in Fig.~\ref{fig:Branch_wavefunctions}, are in agreement with the exact numerical results.

To further investigate the structure of the repulsive to attractive polaron crossover, in Fig.~\ref{fig:transition_near_unitarity} we plot the exact GPe solution profile together with 
the corresponding zero energy Schr\"odinger equation in the vicinity of the first unitary point of the odd solution. To demonstrate the generality of our approach, we present results for both the square-well potential and an exponential potential $\mathcal{U}(y) = -g_e^2e^{-y}$. 
We observe excellent agreement between the exact numerical solution to the GPe and the zero-energy Schr\"odinger solution within the spatial extent of the impurity potential. We note that exactly at unitarity, the tail of the polaronic solution is flat, so it does not give any contribution to the energy in Eq.~\eqref{eq:penegytext}, while the interior of the impurity gives a small non-universal contribution. The criterion of applicability of the weak-coupling regime in Eq.\eqref{eq:dpwc} is $\xi\ll|a|\ll\xi^2/r_c$, and away from the unitarity, the contribution from the tail dominates the energy of the state, determining whether the polaron is repulsive or attractive.

\textit{Conclusion and outlook--}
In this Letter, we studied 
both ground and excited state solutions to the 1D GPe for a static impurity interacting with a Bose gas via a finite-range potential. We showed the appearance of new excited state energy branches as new bound states appear in the potential, and characterized the evolution of such solutions from solitonic to polaronic configurations, and found the analytical expressions for their energies in the universal regimes. We expect that a similar soliton–polaron hybridization can appear in higher dimensions, and the inclusion of the finite-range boson-impurity potential is important for its description. It would also be interesting to investigate how impurities can couple to other topological configurations, such as vortices. 
Since our analysis was based on the mean-field GPe equation for a static impurity, we also aim to investigate how this framework gets modified by the inclusion of quadratic fluctuations on top, and by accounting for a finite impurity mass. 

\vspace{5mm}
\begin{acknowledgments}
This work was supported by the Simons Collaboration on UltraQuantum Matter, which is a grant from the Simons Foundation (651440, VG). NY acknowledges support by AFOSR Grant No.~FA9550-23-1–0598.
\end{acknowledgments}

\bibliography{references}

\end{document}